\documentclass[%
 reprint,
superscriptaddress,
 amsmath,amssymb,
 aps,
pra,
]{revtex4-2}

\usepackage{graphicx}
\usepackage{dcolumn}
\usepackage{bm}
\usepackage{amsmath}

\begin{document}

\preprint{APS/123-QED}

\title{Nematic quantum criticality in an Fe-based superconductor revealed by strain-tuning}

\author{Thanapat Worasaran}
	\thanks{These authors contributed equally to this work}
\affiliation{Stanford Institute for Materials and Energy Science, SLAC National Accelerator Laboratory, \\ 2575 Sand Hill Road, Menlo Park, CA 94025, USA}
\affiliation{Department of Applied Physics and Geballe Laboratory of Advanced Materials, \\ Stanford University, Stanford, CA 94305, USA}
\author{Matthias S. Ikeda}%
	\thanks{These authors contributed equally to this work}
\affiliation{Stanford Institute for Materials and Energy Science, SLAC National Accelerator Laboratory, \\ 2575 Sand Hill Road, Menlo Park, CA 94025, USA}
\affiliation{Department of Applied Physics and Geballe Laboratory of Advanced Materials, \\ Stanford University, Stanford, CA 94305, USA}
\author{Johanna C. Palmstrom}
\affiliation{Stanford Institute for Materials and Energy Science, SLAC National Accelerator Laboratory, \\ 2575 Sand Hill Road, Menlo Park, CA 94025, USA}
\affiliation{Department of Applied Physics and Geballe Laboratory of Advanced Materials, \\ Stanford University, Stanford, CA 94305, USA}

\author{Joshua A. W. Straquadine}
\affiliation{Stanford Institute for Materials and Energy Science, SLAC National Accelerator Laboratory, \\ 2575 Sand Hill Road, Menlo Park, CA 94025, USA}
\affiliation{Department of Applied Physics and Geballe Laboratory of Advanced Materials, \\ Stanford University, Stanford, CA 94305, USA}

\author{Steven A. Kivelson}
\affiliation{Stanford Institute for Materials and Energy Science, SLAC National Accelerator Laboratory, \\ 2575 Sand Hill Road, Menlo Park, CA 94025, USA}
\affiliation{Department of Physics and Geballe Laboratory of Advanced Materials, \\ Stanford University, Stanford, CA 94305, USA}

\author{Ian R. Fisher}
\affiliation{Stanford Institute for Materials and Energy Science, SLAC National Accelerator Laboratory, \\ 2575 Sand Hill Road, Menlo Park, CA 94025, USA}
\affiliation{Department of Applied Physics and Geballe Laboratory of Advanced Materials, \\ Stanford University, Stanford, CA 94305, USA}

\date{\today}

\begin{abstract}
\end{abstract}


\maketitle

{\bf Quantum criticality has been invoked as being essential to the understanding of a wide range of exotic electronic behavior, including heavy Fermion and unconventional superconductivity, but conclusive evidence of quantum critical fluctuations has been elusive in many materials of current interest.
An expected characteristic feature of quantum criticality is power law behavior of thermodynamic quantities as a function of a non-thermal tuning parameter close to the quantum critical point (QCP). 
In the present work, we observe power law behavior of the critical temperature of the coupled nematic/structural phase transition as a function of uniaxial stress in a representative family of Fe-based superconductors. 
Our measurements provide direct evidence of quantum critical nematic fluctuations in this material.
Furthermore, these quantum critical fluctuations are not confined within a narrow regime around the QCP, but extend over a wide range of temperatures and tuning parameters.}

Long range electronic nematic order (defined as electronic order that only breaks point group symmetries) is a ubiquitous feature of iron-based superconductors (see for example references \cite{Fernandes_2012,QimiaoReview} and references therein). For cuprate superconductors, mounting evidence points towards a generic in-plane electronic anisotropy for underdoped compositions, implying the presence of at least a nematic component to an electronic ordered state, and possibly even a vestigial nematic state (see for example references \cite{Nie7980,Mukhopadhyay13249} and references therein).
From a theoretical perspective, several lines of reasoning suggest that nematic fluctuations can provide a pairing interaction \cite{cooperPairing,CuO2Pairing,Lederer4905}, and hence suggest that the presence of nematic order in the phase diagrams of these high temperature superconductors may not be coincidental. 
In particular, nematic fluctuations enhance superconductivity in any symmetry channel \cite{SamEnhancement}, and hence could be a key element for increasing the critical temperature even when the dominant pairing interaction arises from spin fluctuations.
A key open question is whether \emph{quantum critical} nematic fluctuations are present, and if so over how much of the phase diagram. 
The present work directly addresses this latter question for a representative family of Fe-based superconductors, revealing the presence of quantum critical nematic fluctuations via power law variation of the critical temperature of the nematic phase transition with respect to a non-thermal tuning parameter. 

\begin{figure*}
  \includegraphics[width=\linewidth]{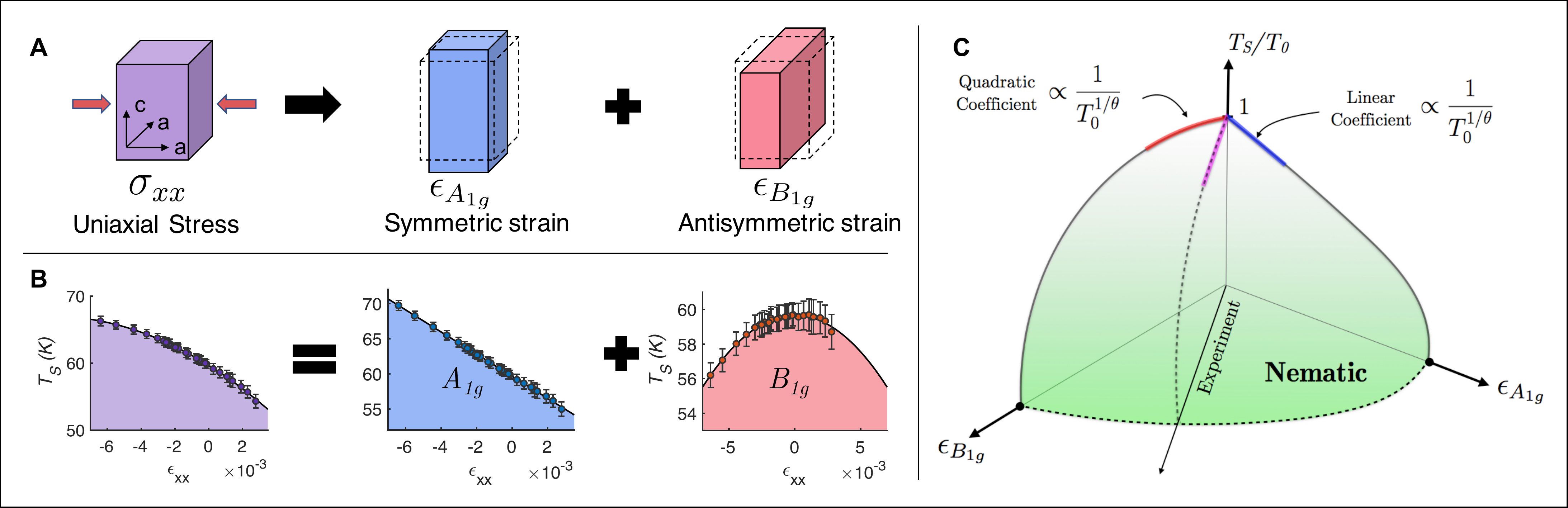}
  \caption{{\bf Consequences of uniaxial stress along [100]} (A) Deformation caused by the uniaxial stress applied along [100] direction can be expressed as the combination of the symmetry preserving strain $\epsilon_{A_{1g}}$, and orthogonal antisymmetric strain $\epsilon_{B_{1g}}$. The ratio of these strain components depends on the elastic moduli of the material. Note that orthogonal antisymmetric $B_{1g}$ strain is defined as $\epsilon_{B_{1g}}=  (\epsilon_{xx}-\epsilon_{yy})/2$, while symmetry preserving $A_{1g}$ strain, in this scenario, is a combination of the in-plane $\epsilon_{A_{1g,1}} = (\epsilon_{xx}-\epsilon_{yy})/2$, and out of plane $\epsilon_{A_{1g,2}}=\epsilon_{zz}$. (B) The variation of coupled nematic/structural phase transition temperature as a function of the measured strain along [100], $\epsilon_{xx}$ for a representative sample with cobalt concentration $x = 4.8\% \pm 0.2\%$. The leftmost graph (purple) shows the (linear + quadratic) variation caused by uniaxial stress. The center graph (blue) shows the linear contribution from $A_{1g}$ strain, and the rightmost graph (red) shows the quadratic contribution from $B_{1g}$ strain. (C) Schematic showing $T_S-\epsilon_{A_{1g}}-\epsilon_{B_{1g}}$ phase diagram. The experimental path lies along the straight line between $\epsilon_{A_{1g}}$ and $\epsilon_{B_{1g}}$ axes. The effects of these two tuning parameters are generally uncorrelated. However, in the presence of strong quantum critical fluctuations, both coefficients are related to the zero strain transition temperature as $1/T_0^{1/\theta}$.} 
  \label{Fig1}  
 \end{figure*}

Direct evidence for a quantum critical regime in the Fe-based superconductors has been limited thus far.
A divergence of the effective mass in BaFe$_2$(As$_{1-x}$P$_x$)$_2$ has been inferred from penetration depth, quantum oscillation, heat capacity, and resistivity measurements \cite{Hashimoto1554,shishido,walmsley,James_transport,shibauchireview}, strongly suggesting the presence of a quantum critical point.
These measurements cannot, however, establish the character of the fluctuating order, in particular whether it is nematic or magnetic.
Measurements of the doping and temperature dependence of the nematic susceptibility for a wide variety of Fe-based superconductors, obtained via elastoresistivity \cite{ChuScience2012,hhscience}, elastic constant measurements \cite{3pointsbending}, Raman scattering \cite{Massat9177,RamanPRL2013,Raman_quadrapole}, and nuclear magnetic resonance \cite{NCurroNMR}, reveal the presence of strong nematic fluctuations and thus are strongly suggestive of the presence of a nematic quantum critical point beneath the superconducting 'dome' in all of these materials.
Recent elastoresistivity measurements in Ba(Fe$_{1-x}$Co$_x$)$_2$As$_2$ for a fine comb of overdoped compositions approaching the critical doping ($x_c$) are consistent with a power law divergence of the nematic susceptibility with respect to $(x-x_c)$, but the temperature dependence is not currently understood in detail, and hence the regime over which quantum critical fluctuations extends is unknown \cite{palmstrom2019comparison}.
Significantly, attempts to observe power law behavior in thermodynamic quantities as a response to a non-thermal tuning parameter upon approaching the putative quantum critical point face daunting challenges.
These challenges are associated either with difficulties in accurately determining the magnitude of the tuning parameters as the material is tuned infinitesimally close to the putative quantum critical point (this is the case for chemical substitution) or in obtaining sufficient fine-tuning of the tuning parameter in that regime (which can be the case for hydrostatic pressure, another common tuning parameter). 
Here we bypass these difficulties by using symmetric strain ($\epsilon_{A_{1g}}$) and orthogonal antisymmetric strain ($\epsilon_{B_{1g}}$) induced by in-plane uniaxial stress as essentially-continuously-variable tuning parameters. 
By doing so, we are able to show for underdoped compositions of the representative Fe-based superconductors Ba(Fe$_{1-x}$Co$_x$)$_2$As$_2$ that a single power law governs the variation rate of the critical temperature of the coupled nematic/structural phase transition ($T_S$) with respect to \emph{both} of these symmetry-inequivalent tuning parameters.
This provides the first direct evidence of quantum critical power law behavior in this representative material.
Moreover, the technique we demonstrate establishes a novel framework to observe quantum criticality in other materials in this class and beyond.

Strain induced by external stresses only breaks point symmetries, and consequently has a special role to play in the study of electronic nematicity.
In recent years, considerable progress has been made exploiting the fact that strain with the same symmetry as the nematic order acts as an effective conjugate field due to the bilinear coupling  between the strain and the nematic order parameter (see for example references \cite{ChuScience2012,Max_ER,BOHMER201690}).
Here, we explore how strains of \emph{different} symmetries couple to Ising nematic order and determine the shape of the phase boundary in temperature-strain space.
These ideas are not specific to nematicity (since they do not rely on bilinear coupling) and could be applied to an even wider set of strain tuned phase transitions.

The representative materials, Ba(Fe$_{1-x}$Co$_x$)$_2$As$_2$, belong to the $D_{4h}$ point group and undergo a phase transition to an electronic nematic state with $B_{2g}$ symmetry.
Due to the coupling between the electronic and lattice degrees of freedom, the phase is also characterized by a spontaneous $B_{2g}$ strain, $\epsilon_{B_{2g}} \equiv \epsilon_{xy}$.
We consider the effects of strains belonging to two other irreducible representations of the point group, namely the symmetry preserving strain $\epsilon_{A_{1g}}$, and the \emph{orthogonal} antisymmetric strain $\epsilon_{B_{1g}}$ (Fig.\ref{Fig1}A center and right panels).
Here $\epsilon_{A_{1g}}$ is a combination of in-plane ($\epsilon_{A_{1g,1}} \equiv (\epsilon_{xx} + \epsilon_{yy})/2$) and out-of-plane ($\epsilon_{A_{1g,2}} \equiv \epsilon_{zz}$) symmetric strains, and $\epsilon_{B_{1g}}$ is defined as $\epsilon_{B_{1g}} \equiv (\epsilon_{xx} - \epsilon_{yy})/2$ (see Appendix \ref{Apndx:Decomposing}). 
Note here that the coordinates are defined in the two-iron unit cell, i.e. $\epsilon_{xx}$ and $\epsilon_{xy}$ are the deformation along Fe-As bond and Fe-Fe bond, corresponding to [100] and [110] crystallographic directions respectively.
The tuning effect of ${A_{1g}}$ and ${B_{1g}}$ strains are different, as has been shown previously \cite{mitw}.
In the small strain regime (which applies to this work), ${A_{1g}}$ strain tunes the critical temperature $T_S$ linearly to leading order.
The orthogonal antisymmetric strain $\epsilon_{B_{1g}}$, however, can vary the critical temperature $T_S$ only quadratically to leading order; linear variation with respect to $\epsilon_{B_{1g}}$ is prohibited by symmetry \cite{Tsb1g}.
Hence, up to quadratic order, the variation of the critical temperature $T_S$ in the presence of $\epsilon_{A_{1g}}$ and $\epsilon_{B_{1g}}$ strains is
\begin{equation}
	T_S(\epsilon_{A_{1g}}, \epsilon_{B_{1g}}) = T_0+a\epsilon_{A_{1g}} + a'\epsilon_{A_{1g}}^2+b\epsilon_{B_{1g}}^2
\end{equation}
where $T_0 \equiv T_S(0, 0)$ is the free standing critical temperature, and $a$, $a'$ and $b$ are the rate of variation of the critical temperature due to $\epsilon_{A_{1g}}$ and $\epsilon_{B_{1g}}^2$ respectively. 
Of particular relevance to the present work, measurements performed under hydrostatic pressure conditions reveal that the coefficient $a'$ is negligibly small for all compositions studied (see Appendix \ref{Apndx:Decomposing}).
Hence, the tuning effect of $\epsilon_{A_{1g}}$ and $\epsilon_{B_{1g}}$ can be unambiguously disentangled, even in the presence of both symmetry strains.
Since $A_{1g}$ and $B_{1g}$ strains belong to different irreducible representations, the coefficients $a$ and $b$ are ordinarily anticipated to be completely independent - i.e. unrelated by any symmetry operations. 
As we will demonstrate below, this is no longer the case in the presence of strong quantum critical fluctuations.

Uniaxial stress was applied to Ba(Fe$_{1-x}$Co$_x$)$_2$As$_2$ single crystals using a commercially available strain cell (CS100, {\it Razorbill Instrument}) and varied in-situ in an almost continuous fashion.
Bar-shaped single crystalline samples (with typical dimensions of $2000 \times 500 \times 35 \mu m$, cut along [100] direction) are glued onto two mounting plates, that can be pushed/pulled by varying the voltage to sets of lead-zirconium-titanate (PZT) piezoelectric stacks (see Appendix \ref{Apndx:Method}).
The strain cell is designed to compensate for thermal expansion of the PZT  \cite{HicksMechanism}, and since the differential thermal expansion of the cell body and the sample is negligible, the strain is almost perfectly temperature independent at a fixed voltage \cite{mitw}.
Strain along the [100] crystallographic direction, $\epsilon_{xx}$, can be inferred from the change in the mounting plate separation by measuring the change in capacitance of the capacitive sensor inside the cell body using a capacitance bridge by {\it Andeen-Hagerling} (AH2550A).
Through finite element simulation, the strain relaxation through the glue can be estimated, and the strain experienced by the sample is found to be approximately $70\%$ of the measured strain \cite{mitw}.
The strains $\epsilon_{A_{1g}}$ and $\epsilon_{B_{1g}}$ are related to $\epsilon_{xx}$ via the elastic moduli $c_{ijkl}$ of the samples. 
It can be shown (see Appendix \ref{Apndx:Decomposing}) that a variation in $c_{ijkl}$ within the composition and temperature range investigated here has no effect on the conclusions of this work.
The critical temperature $T_S$ is determined using AC elastoresistivity \cite{ACER} outlined in Appendix \ref{Apndx:Method}, as well as in some cases AC elastocaloric effect described in detail elsewhere \cite{IkedaEC}.

Fig.\ref{Fig1}A illustrates schematically how the deformation due to [100] uniaxial stress can be expressed as a linear combination of the symmetric and antisymmetric strains, $\epsilon_{A_{1g}}$ and $\epsilon_{B_{1g}}$.
Using the fact that $\epsilon_{A_{1g}}$ and $\epsilon_{B_{1g}}$ are both linearly proportional to $\epsilon_{xx}$ (see Appendix \ref{Apndx:Decomposing}) , and that the coefficient $a'$ is vanishingly small, the variation of $T_S$ due to [100] uniaxial stress is
\begin{equation}
T_S(\epsilon_{xx}) = T_0 + \alpha \epsilon_{xx} + \beta \epsilon_{xx}^2,
\label{eq:Tsstrain}
\end{equation}
where $\alpha = dT_S/d\epsilon_{xx} \propto \partial T_S/\partial \epsilon_{A_{1g}} $ and $\beta = 1/2 \times d^2T_S/d\epsilon_{xx}^2 \propto  \partial^2 T_S/\partial \epsilon_{B_{1g}}^2$. 
Fig.\ref{Fig1}B shows representative data for a sample with composition $x = 4.8 \% \pm 0.2 \%$ revealing the linear and quadratic behavior.

\begin{figure}
  \includegraphics[width=\linewidth]{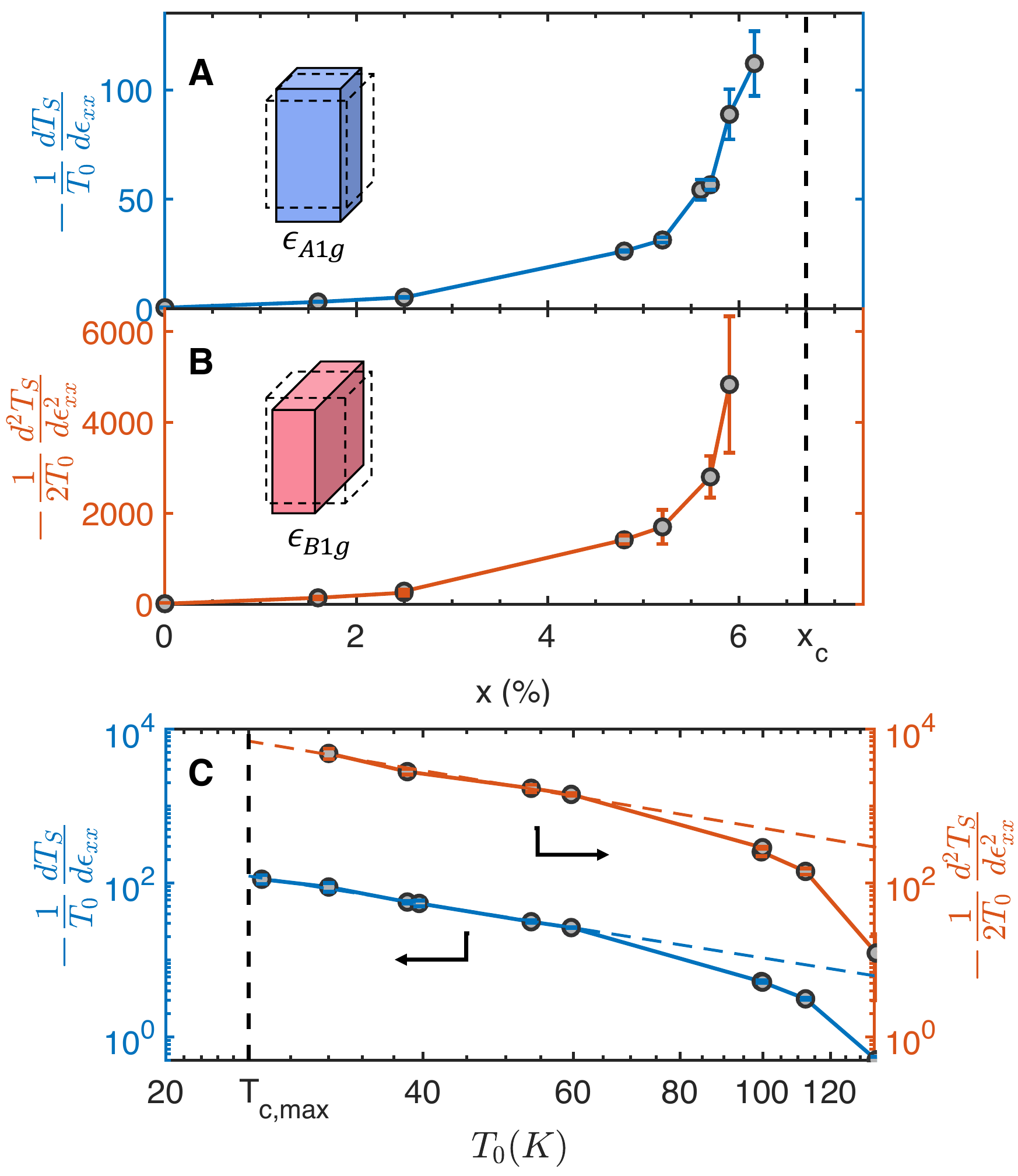}
  \caption{{\bf Power law divergence of $\mathbf{dT_S/d\epsilon_{xx}}$ and $\mathbf{d^2T_s/d\epsilon_{xx}^2}$ as a function of cobalt concentration x in $\mathbf{Ba(Fe_{1-x}Co_{x})_2As_2}$.} (A) The linear coefficient of the variation of $T_S$ caused by symmetric ($A_{1g}$) strain, and (B) the quadratic coefficient caused by antisymmetric ($B_{1g}$) strain, each case normalized by the zero strain critical temperature $T_0$. The insets illustrate the strain irreducible representations. Both the linear and the quadratic coefficient diverge as the cobalt concentration approaches the critical doping of $x_c \sim 6.7\%$ (vertical dashed line). (c) The same normalized coefficients as a function of zero-strain critical temperature $T_0$ shown in a log-log scale. The linear behavior seen in this plot at low temperatures indicates similar power law behavior for both symmetry channels. The slope of the linear regimes is equal to the inverse critical exponent $1/\theta$, which is found to be $1/\theta= 1.75 \pm 0.2$ and $1/\theta=1.87 \pm 0.7$ for $A_{1g}$ and $B_{1g}$ strain respectively. Remarkably these values agree within a standard deviation. Note that extracting statistically significant quadratic coefficients requires data extending to relatively large strains. For two experiments (two extra data points in the blue traces), this strain regime could not be reached before the mechanical failure of the sample, thus only the linear coefficients are shown.}
  \label{FigResult}
\end{figure}

Further investigation of the coefficients $\alpha$ and $\beta$ within the doping series of Ba(Fe$_{1-x}$Co$_x$)$_2$As$_2$ reveals a striking result.
Fig.\ref{FigResult}A shows the normalized coefficients $\alpha/T_0$ and $\beta/T_0$ as a function of cobalt concentration $x$.
Both normalized coefficients grow monotonically and appear to diverge as the cobalt concentration approaches the critical doping, $x_c$. 
Here, $x_c$ is defined in the absence of superconductivity (see Fig.\ref{FigQC}B), and was recently measured to be $x_c \sim 6.7 \pm 0.2 \%$  using high magnetic fields to suppress the superconducting phase \cite{palmstrom2019comparison}.

\begin{figure*}
  \includegraphics[width=1\linewidth]{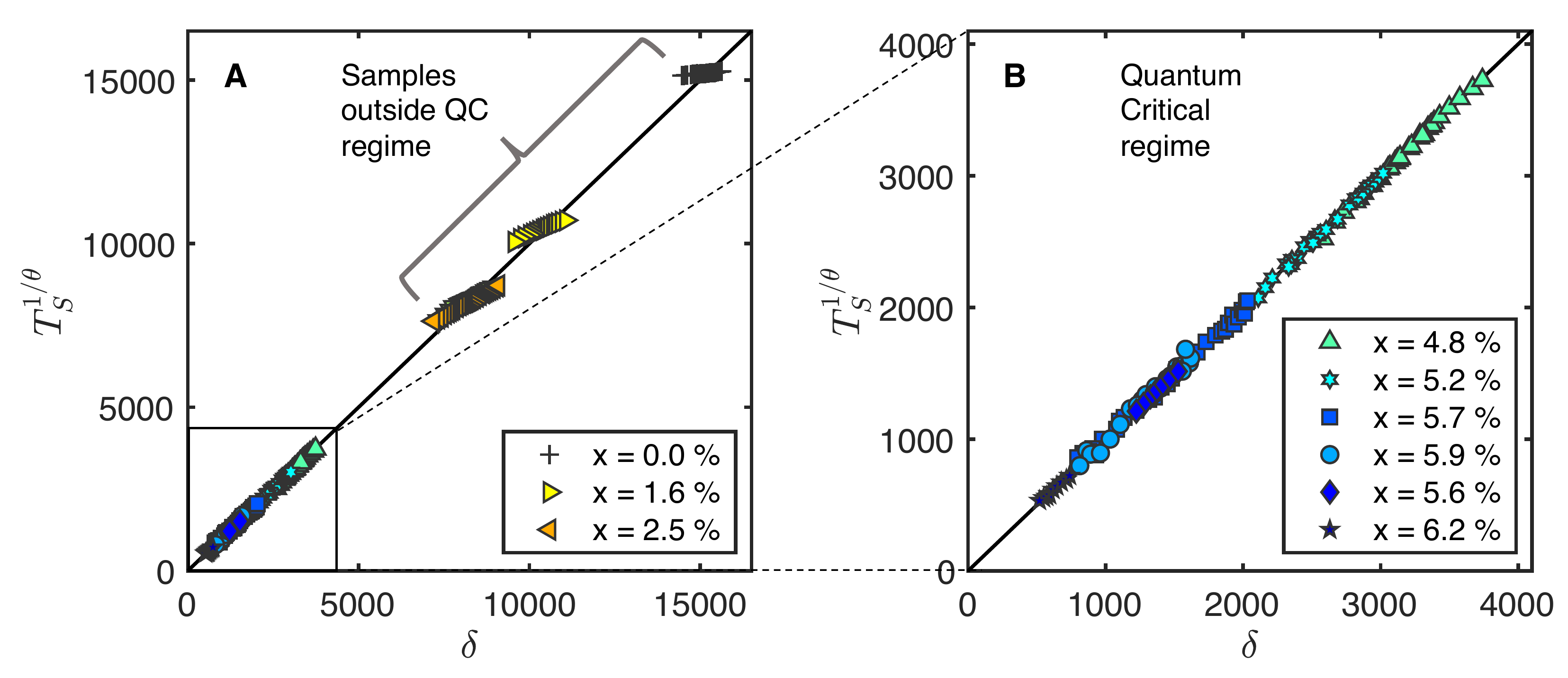}
  \caption{{\bf Scaling collapse of non-thermal tuning parameters.} The scaling collapse described in Eq.(5) is illustrated in (A) where $T_S^{1/\theta}$ is plotted versus $\delta$. The best fit value $\theta = 0.52 \pm 0.02$ is extracted using the method explained in Appendix \ref{Apndx:Scale}. Symbols represent samples with differing dopings. The data is expected to collapse onto a single line in the quantum critical regime which is observed for samples within a range of Co concentrations between $x= 4.8$\% and 6.2\%. This regime is magnified in the plot in panel (B) which shows a remarkable collapse over almost a decade of $\delta$.}
  \label{FigScaleCollapse}
 \end{figure*}

In what follows we argue that a power law describing the divergence of $\alpha/T_0$ and $\beta/T_0$ is a hallmark for quantum critical fluctuations.
We start from the ansatz:
\begin{equation}
T_S \sim (g_C-g)^{\theta},
\label{eq:critscaling}
\end{equation}
which suggests that within a regime dominated by quantum critical fluctuations, the finite temperature phase transition is governed by a power law of a single (relevant) non-thermal tuning parameter, $g$ \cite{sachdev}.
Here, $g_C$ is the value of $g$ at the QCP, and $\theta$ is the critical exponent governing the phase boundary in the immediate vicinity of the QCP \cite{nuz}.
In the context of this work, $g$ can be varied by varying either $\epsilon_{A_{1g}}$ or $\epsilon^2_{B_{1g}}$ \cite{geb1gdquare}, and we reiterate that these are distinct tuning parameters belonging to orthogonal symmetry channels.

Taylor expanding Eq.(3) in a small strain region around zero strain, and substituting $g_C \sim T_0^{1/\theta}$, we arrive at a simple expression for the leading order effect of the tuning parameter $g$ on $T_S$ in the quantum critical regime: 
\begin{equation}
	\frac{T_S}{T_0} =  1 - \frac{C\theta}{T_0^{1/\theta}} g+O(g^2),
	\label{eq:scaleslope}
\end{equation}
where $C$ is a constant. 
Substituting $\epsilon_{A_{1g}}$ or $\epsilon^2_{B_{1g}}$ for $g$ gives $\alpha/T_0 \sim -T_0^{-1/\theta}$ and $\beta/T_0 \sim -T_0^{-1/\theta}$ respectively.
This is the power law that governs the morphology of the strain-tuned phase diagram in the presence of quantum critical fluctuations.
Notice in particular that in the quantum critical regime, the same power law governs the behavior of \emph{both} symmetry channels, even though these two strains belong to distinct irreducible representations of the point group (Fig.\ref{Fig1}C). 
Notice also how the composition $x$ does not enter into Eq.(4). 
All that is needed is a measurement of $T_0$ and of the rate of suppression of $T_S$ with respect to $\epsilon_{A_{1g}}$ and $\epsilon_{B_{1g}}^2$. 
If the material is in the quantum critical regime, tuned by $x$, then these two coefficients will follow the same power law behavior. 

\begin{figure}
  \includegraphics[width=\linewidth]{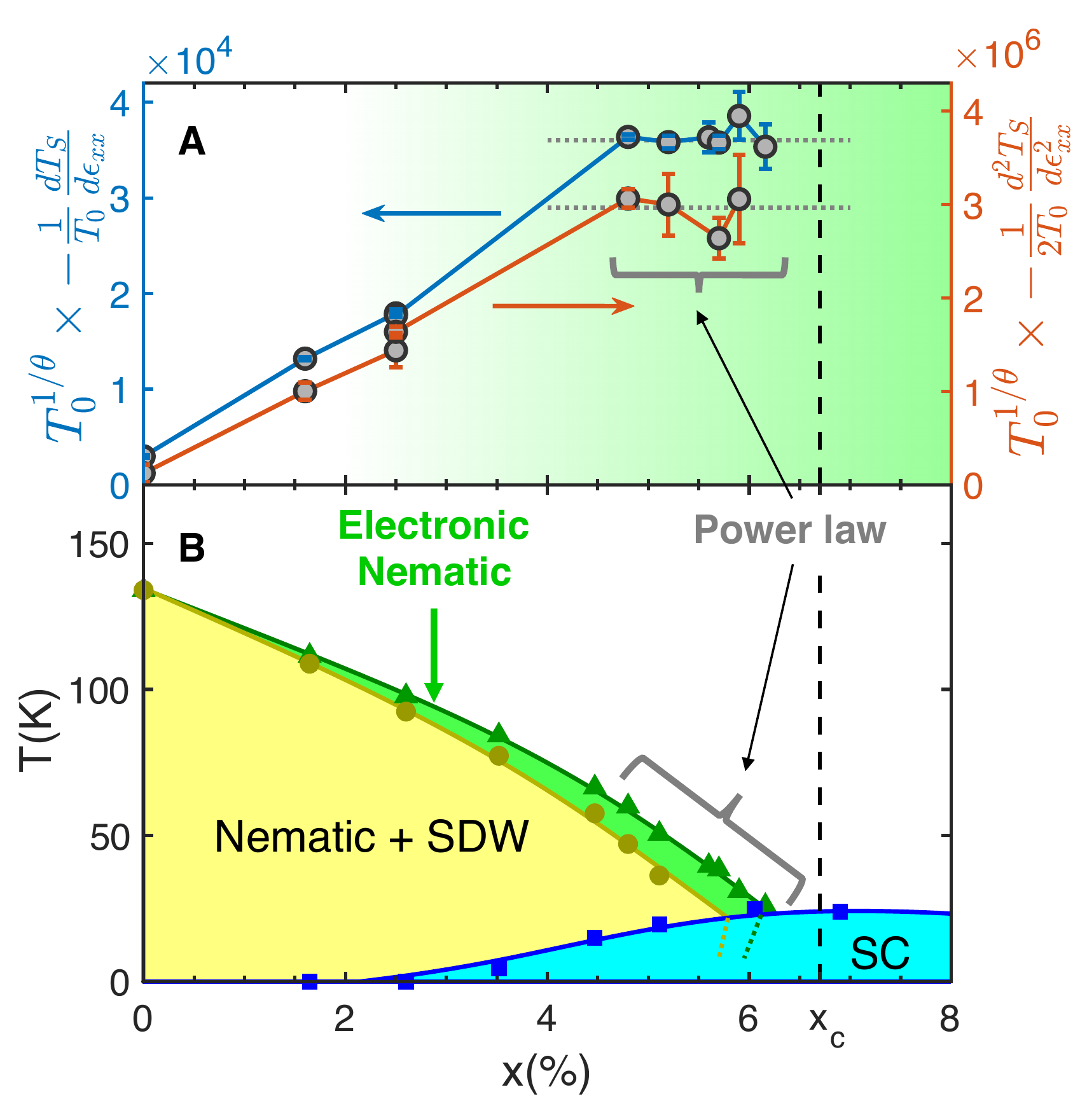}
  \caption{{\bf Quantum criticality in $\mathbf{Ba(Fe_{1-x}Co_{x})_2As_2}$ phase diagram.} (A) The normalized linear $\alpha/T_0 $ (left axis) and quadratic coefficients $\beta/T_0 $ (right axis), scaled with the expected zero-strain critical temperature dependent $T_0^{-1/\theta}$. These quantities are expected to be constant in the quantum critical regime, which spans the higher dopings, as suggested by the guide dotted line. The green shading represent the ratio of these scaled coefficients with their saturated value. (B) The phase diagram of Ba(Fe$_{1-x}$Co$_x$)$_2$As$_2$ \cite{JH_phase}. Power law behavior is observed for materials with a cobalt concentration between $x=4.8\% \pm 0.2\%$ all the way to $x=6.2\% \pm 0.2\%$, above which $T_S$ is unobservable due to the presence of superconductivity. This power law behavior signifies the presence of strong nematic fluctuation of quantum origin due to the presence of the nematic QCP at $x_c$. Note that the dotted lines in (B) denotes the phase transition lines which presumably backbend inside the superconducting state \cite{AmesBackBending}.}
  \label{FigQC}
\end{figure}

The distinct advantages of this perspective lies in (i) the fine tunability of $\epsilon_{xx}$, allowing for the accurate determination of $dT_S/d\epsilon_{xx}$ and $d^2T_S/d\epsilon_{xx}^2$; (ii) the simultaneous determination of the effect of \emph{two} tuning parameters (symmetric and antisymmetric strains) within \emph{one} single experiment; (iii) the circumvention of the large uncertainties in determining the chemical composition - i.e. the determination of the chemical composition is unnecessary; and (iv) our approach completely bypasses the need to determine the critical value of the tuning parameter $g_C$ (with respect to composition or strain), because the analysis investigates the variation of $T_S$ with respect to the tuning parameter in the limit of small strains (this is especially useful in the present context due to the presence of the superconducting dome, but more generally it eliminates all errors one would ordinarily incur due to considering decades of variation in the quantity $\delta = g-g_C$ when both $g$ and $g_C$ suffer from large uncertainty).

The power law behavior in $\alpha/T_0$ and $\beta/T_0$ is visualized best as a log-log plot, for which a straight line with a slope of $1/\theta$ is expected for both coefficients.
This is precisely what is observed as is shown in Fig.\ref{FigResult}B. As the  zero-strain critical temperature ($T_0$) decreases (tuned by composition, $x$), there is an apparent tendency for the strain-coefficients towards a linear behavior in the log-log plot, indicated by the dashed lines.
The cross-over regime appears at a similar value of $T_0$ for both $\alpha/T_0$ and $\beta/T_0$. 
Moreover, the slopes of the best-fits which represent the inverse critical exponents appear to be similar in both cases, with the value $1/\theta = 1.75 \pm 0.2$ for $\alpha/T_0$ , and $1/\theta = 1.87 \pm 0.7$ for $\beta/T_0$, agreeing within experimental error.
Averaging the two exponents obtained from the temperature dependence of the linear and quadratic coefficients, we obtain $\theta = 0.56 \pm 0.07$. 

Equivalently, this can also be illustrated by a scaling collapse with respect to the non-thermal tuning parameters. As a function of the 3 tuning parameters, $x$, $\epsilon_{A_{1g}}$ and $\epsilon_{B_{1g}}$, the critical temperature $T_S$ varies as
\begin{multline}
	T_S(x,\epsilon_{A_{1g}},\epsilon_{B_{1g}})  = \delta ^\theta \\ = [T_0(x)^{1/\theta} + A_1\epsilon_{A_{1g}} + A_2\epsilon_{B_{1g}}^2]^\theta
\end{multline}
in the quantum critical regime. Here $T_0(x) \equiv T_S(x,0,0)$, $A_1$ and $A_2$ are constants. 
With appropriate fit values of $A_1$, $A_2$, and $\theta$, the data points should collapse into a single line inside the quantum critical regime. 
This is exhibited in Fig.\ref{FigScaleCollapse} where a linear relationship of $\delta$ and $T_S^{1/\theta}$ can be seen in strain-tuned samples with $T_S$ ranging from $\sim$25K up to $\sim70$K. Using this method, $\theta = 0.52 \pm 0.02 $ provides the best collapse, agreeing with the value obtained from the power law analysis (for more details, see Appendix \ref{Apndx:Scale}).
This value of $\theta$ implies that the $x$ derivative of $T_S$ diverges as $x$ approaches $x_c$, which is consistent with the shape of the phase diagram (Fig.\ref{FigQC}B). 
Presumably this value of $\theta$ relates to universal exponents of a universality class appropriate for metallic Ising nematic systems with disorder, and as such provides a test for future theoretical treatments.
We note in passing that the measured value is equal to the mean field value ($\theta = 0.5$) within experimental error, a value which has been predicted based on perturbative renormalization of the tuning parameter. \cite{QCElasticity,sachdev}.

Ideally, power law behavior is established over decades of a tuning parameter. 
In the present case, however, the presence of superconductivity at $T_{c,max} \sim 25K$ limits the accessible parameter range (chemical doping$/T_0$) for our investigation.
Nevertheless, the observation of such closely similar behavior for the two fully-independent tuning parameters $\epsilon_{A_{1g}}$ and $\epsilon_{B_{1g}}$, and the scaling collapse shown in Fig.\ref{FigScaleCollapse}, provide compelling evidence that the apparent power law behavior is driven by critical fluctuations.

The power law variation of $\alpha/T_0 \sim -1/T^{1/\theta}$ and $\beta/T_0 \sim -1/T^{1/\theta}$ implies that the normalized quantities, $T^{1/\theta} \times \alpha/T_0$ and $T^{1/\theta} \times \beta/T_0$, should be constant so long as the material is in the quantum critical regime. 
This behavior is verified in Fig.\ref{FigQC}A, which shows the variation of the normalized quantities as a function of composition, $x$.  
This figure makes clear two very important points. 
Firstly, the composition $x$ does not need to be accurately determined to deduce that the material is in the quantum critical regime - it is sufficient solely that the normalized quantities have constant values; the absolute value of the $x$-coordinates in Fig.\ref{FigQC}A does not matter in reaching this conclusion.
Secondly, the regime of power law exists over a remarkably wide regime of composition and temperature (gray brackets in Fig.\ref{FigQC}A and Fig.\ref{FigQC}B), extending over the majority of the superconducting dome for these underdoped compositions. The immediate implications for superconductivity are unknown. 
However, these experimental results empirically establish that the superconductor is not only born out of a metal that hosts strong nematic fluctuations (as has been previously inferred from a wide variety of measurements \cite{hhscience,ChuScience2012,Massat9177,RamanPRL2013,Raman_quadrapole}), but more specifically is born from a metal that exhibits \emph{quantum critical} nematic fluctuations. 
Put another way, for a wide range of compositions in the $x-T$ plane, the material 'knows' how far it is from the nematic QCP tuned by chemical composition, and this approximately correlates with the range of compositions over which the material superconducts.

\section*{ACKNOWLEDGEMENTS}
The authors thank S. Raghu and P. Walmsley for many insightful discussions.
This work was supported by the Department of Energy, Office of Basic Energy Sciences, under Contract No. DEAC02-76SF00515. M.S.I. was supported in part by the Gordon and Betty Moore Foundations EPiQS Initiative through Grant No.GBMF4414. J.C.P. was supported by a Gabilan Stanford Graduate Fellowship and a Stanford Lieberman Fellowship. J.A.W.S. acknowledges support as an ABB Stanford Graduate Fellow.

\appendix

\section{\label{Apndx:Method} Experimental methods}
\subsection{Straining method}
\subsubsection{Uniaxial stress experiment}

\begin{figure*}
  \includegraphics[width=1\linewidth]{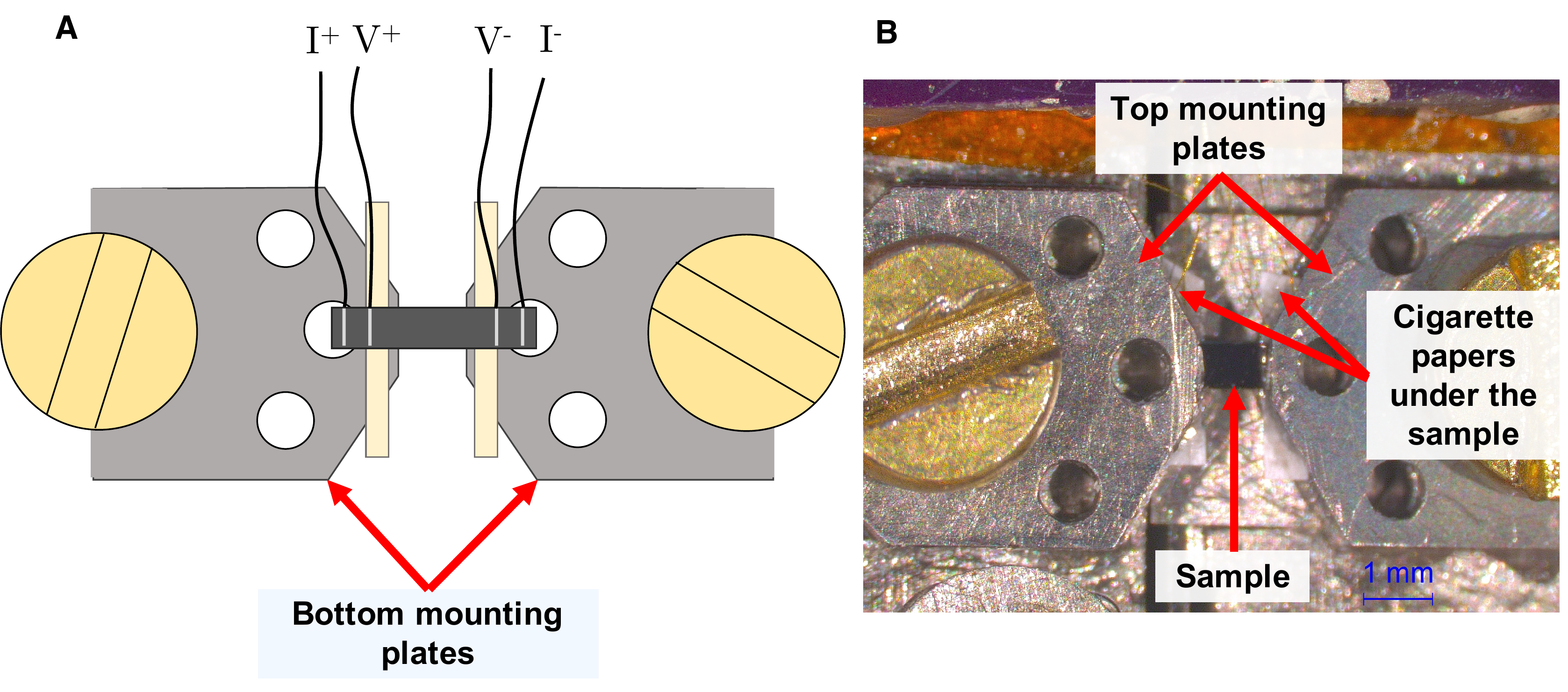}
  \caption{{\bf Strain cell setup.} (A) Schematic showing the assembly procedure. The sample is glued on top of bottom mounting plates and is electrically separated by thin cigarette papers. (B) shows the final arrangement where the top mounting plates are glued on top of the sample and affixed to the cell by screws to provide uniform strain along the $z$-direction. }
  \label{FigExptSetup}
\end{figure*}

The uniaxial stress experiment is carried out in a commercially available strain device ((CS100, {\it Razorbill Instrument}). 
The device is constructed using 3 piezoelectric stacks (PZT) based on the design by Hicks {\it et al.} \cite{HicksMechanism}. 
This design enables the full utilization of the PZT's expansions - yielding up to 1.5\% tensile and compressive strain, which in our case exceeds the yield strength of the samples.
Furthermore, the design compensates for the thermal expansion of the PZT stacks.
The only other source of any eventual temperature dependence of strain at a fixed voltage is the difference in thermal expansion of the sample and the cell body.
In this case it is found that the thermal expansion of titanium and Ba(Fe$_{1-x}$Co$_x$)$_2$As$_2$ are matched at all temperatures \cite{mitw}. 
This means the setup will yield constant strain when sweeping temperature at fixed voltage.

Samples are glued on top of the lower mounting plates with Devcon 2-ton epoxy.
Cigarette paper can be inserted in between to separate the sample from the mounting plates electrically.
Top mounting plates are glued to the samples and are screwed on top of the lower mounting plates separated by washers of fixed thickness to obtain a uniform stress across the sample's cross section.
Strain is inferred from the measured change in separation of the mounting plates and the initial separation of the mounting plates at room temperature. 
The length change is measured using a capacitance sensor inside the strain cell body.
The capacitance is sampled using an {\it Andeen-Hagerling} AH2550A capacitance bridge.
Finite element simulation shows that strain is relaxed through the glue layer and the strain experienced by the sample is reduced from measured strain by a factor of $0.7 \pm 0.07$ \cite{mitw}.
The setup is illustrated in Fig.\ref{FigExptSetup}

To measure the critical temperature, AC elastoresistivity \cite{ACER} (outlined in the next section) is performed using a standard 4-point contact method.
The temperature of the cell is measured using Cernox CX-150 temperature sensor attached to the strain device's body and sampled using Lakeshore 336.
The temperature sweep rate is set to 0.5K/minute for all measurements yielding a temperature lag of the sample with respect to the cell body of around 0.1K.

\subsubsection{Hydrostatic pressure experiment}

Hydrostatic pressure experiments were performed using a {\it Quantum Design} HPC-30 Cu-Be based self-clamping pressure cell. 
Although this version is no longer commercially available, information on the very similar updated version, HPC-33, can be found on the {\it Quantum Design} website. 
Hydrostatic pressure up to $\sim$2.5 GPa is applied using a hydraulic press and Daphne Oil 7373 is used as a pressure transfer medium. 
The freezing point of the Daphne oil is always below room temperature for pressures less than 2 GPa, which ensures a high degree of hydrostaticity throughout the pressure range \cite{Solidification}. 
Pressure measurements were performed by probing the superconducting transition temperature of a lead manometer \cite{SMITH196953}. 
In addition, the temperature dependence of the hydrostatic pressure within the HPC-30 pressure cell was determined by calibration measurements using both a lead and a manganin manometer \cite{PVariation}. 
Below 100 K the hydrostatic pressure was found to be almost independent of temperature.

\begin{figure*}
  \includegraphics[width=1\linewidth]{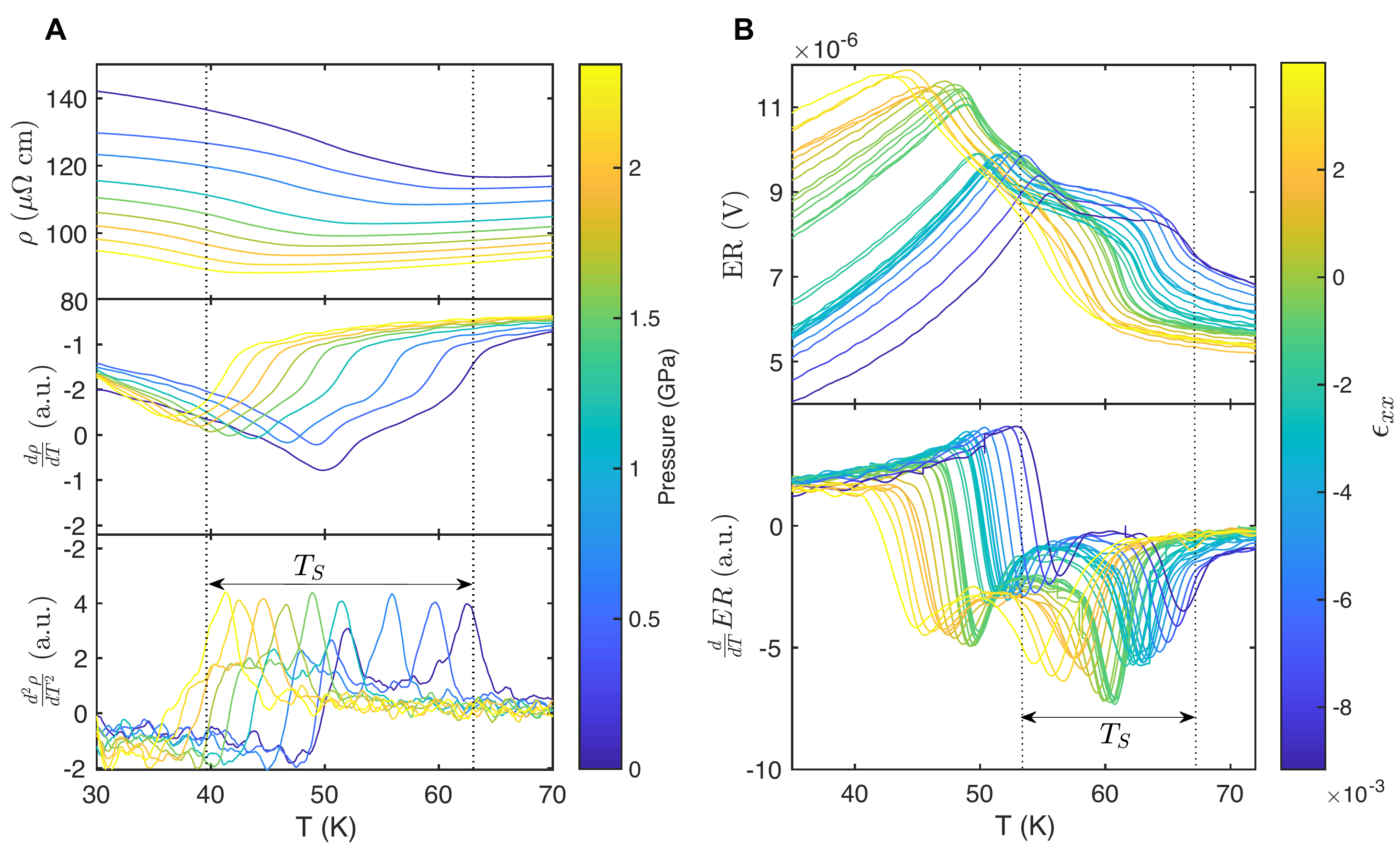}
  \caption{{\bf Measurement of the critical temperature $\mathbf{T_S}$.} (A) Resistivity traces and their temperature derivatives for a prototypical $x = 4.8\% \pm 0.2\%$ under various hydrostatic pressures. The step like anomalies in $d\rho/dT$ found at high temperature of each trace signify the nematic phase transition. $T_S$ is then located at the peak of $d^2\rho/dT^2$. (B) Side band traces and their temperature derivatives for a prototypical $x = 4.8\% \pm 0.2\%$ under various uniaxial stresses. The top panel shows the side band signal which is directly proportional to the elastoresistivity (ER). The nematic phase transition occurs at the step-like anomalies in the ER traces. $T_S$ is thus located at the peaks of $d(ER)/dT$.}
  \label{FigTs}
\end{figure*}

\subsection{Measuring the critical temperature}
\subsubsection{Resistivity as a probe to measure the critical temperature, appropriate for hydrostatic pressure  experiment.}
Since the nematic phase transition behaves like a mean field second order phase transition, a step-like anomaly in the heat capacity signal can be observed at the critical temperature $T_S$ \cite{JH_phase}. 
This anomaly can also be measured through resistivity probe: 
for a continuous transition into a magnetically ordered state \cite{FisherLanger}, scattering from critical degrees of freedom and their energy density are closely related. 
It has been shown \cite{ACER} that this relation is not limited to continuous magnetic transitions but also holds for the nematic phase transition in iron pnictides. 
The heat capacity anomaly around the nematic phase transition leaves an imprint in the temperature derivative of the electrical resistivity.
The critical temperature $T_S$ is extracted from a peak in the second derivative of the electrical resistivity.
As illustrated in Fig.\ref{FigTs}A, step-like anomalies appear in $d\rho/dT$, which translate to a peak in $d^2\rho/dT^2$.

\subsubsection{Elastoresistivity as a probe to measure the critical temperature, appropriate for uniaxial stress experiment.}
Elastoresistivity has recently been introduced as a method \cite{ACER} to detect features of continuous phase transitions. Since the accurate detection of $T_S$ from elastoresistivity is key to this work, below we summarize the ideas presented in \cite{ACER}.

The resistivity close to a continuous phase transition is a function of the reduced temperature  $\rho =\rho(T-T_S)$ \cite{FisherLanger}, and since applying tuning strain changes the critical temperature, $T_S = T_S(\epsilon)$ (see Appendix \ref{Apndx:Decomposing}), the resistivity can be expressed as
\begin{equation*}
	\rho(T,\epsilon) =\rho(T-T_S(\epsilon)).
\end{equation*}
Therefore,
\begin{equation*}
	\frac{\partial}{\partial t}\rho(T-T_S(\epsilon)) = -\frac{\partial}{\partial \epsilon}\rho(T-T_S(\epsilon))\times 1/(\frac{dT_S}{d\epsilon}),
\end{equation*}
and hence,
\begin{equation*}
	\frac{\partial}{\partial t}\rho(T, \epsilon) \propto \frac{\partial}{\partial \epsilon}\rho(T,\epsilon).
\end{equation*}
This implies that measuring the elastoresistivity (i.e. change in resistivity as a response to strain) is equivalent to measuring the temperature derivative of the resistivity and can therefore be used to determine $T_S$.

\subsubsection*{AC Elastoresistivity}

To measure the elastoresistivity, we employ the newly developed AC elastoresistivity technique outlined in \cite{ACER}, which allows us to measure this quantity quickly and with high accuracy. 

Slow oscillating strain ($\sim$ 3Hz) is applied on top of DC offset strain by applying a corresponding voltage signal to the outer PZT stacks.
The resistance of the sample as a function of time can be expressed as 
\begin{equation*}
	R(\epsilon_0+\delta(t)) = R(\epsilon_0)+\left. \frac{dR}{d\epsilon}\right| _{\epsilon=\epsilon_0} \times \delta_0 \sin(\omega_s t).
\end{equation*}
Here, $\delta(t)$ is a small oscillating strain, and $\omega_s$ is its frequency.
AC current of ~$\sim$100Hz is passed through the sample resulting in an AC voltage with slowly varying amplitude which can be expressed as
\begin{multline*}
	V = I(t)R(t) = I_0\sin(\omega_ct)\times R(\epsilon_0) \\ + \frac{I_0}{2} \left. \frac{dR}{d\epsilon}\right| _{\epsilon=\epsilon_0} \delta_0 \times (\cos(\omega_c - \omega_s)t-\cos(\omega_c+\omega_s)t) 
\end{multline*}
where $\omega_c$ is the carrier frequency and the the last term is obtained from trigonometry product identity.
The quantity $dR/d\epsilon$ is measured by lock-in detection of the side band (of either frequency  $\omega_c \pm \omega_s$) using the dual mode of a Stanford Research SR860 lock-in amplifier.
The critical temperature can then be identified by a step-like anomaly in this side band signal (Fig.\ref{FigTs}B).

This technique is particularly useful in uniaxial stress experiments on samples with higher cobalt concentration.
For these materials, the signature of the phase transition is smeared by disorder and strain anisotropy, and thus a superior signal-to-noise ratio of the technique reduces the uncertainty in $T_S$.

\begin{figure}
  \includegraphics[width=0.7\linewidth]{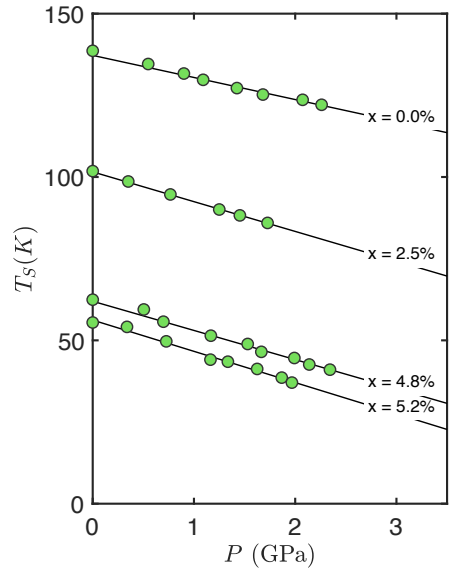}
  \caption{{\bf Response of $T_S$ to applied hydrostatic pressure for various compositions of $\mathbf{Ba(Fe_{1-x}Co_x)_2As_2}$}. The green circles represent the measured critical temperature $T_S$ as a function of applied hydrostatic pressure $P$. The black lines represent the linear fit to the data. It can be seen that in all compositions, the variation is linear. Thus the quadratic variation in uniaxial stress experiment (Fig.\ref{FigUni}) is exclusively caused by the orthogonal antisymmetric $B_{1g}$ strain.}
  \label{FigHyd}
\end{figure}

\begin{figure*}
  \includegraphics[width=1\linewidth]{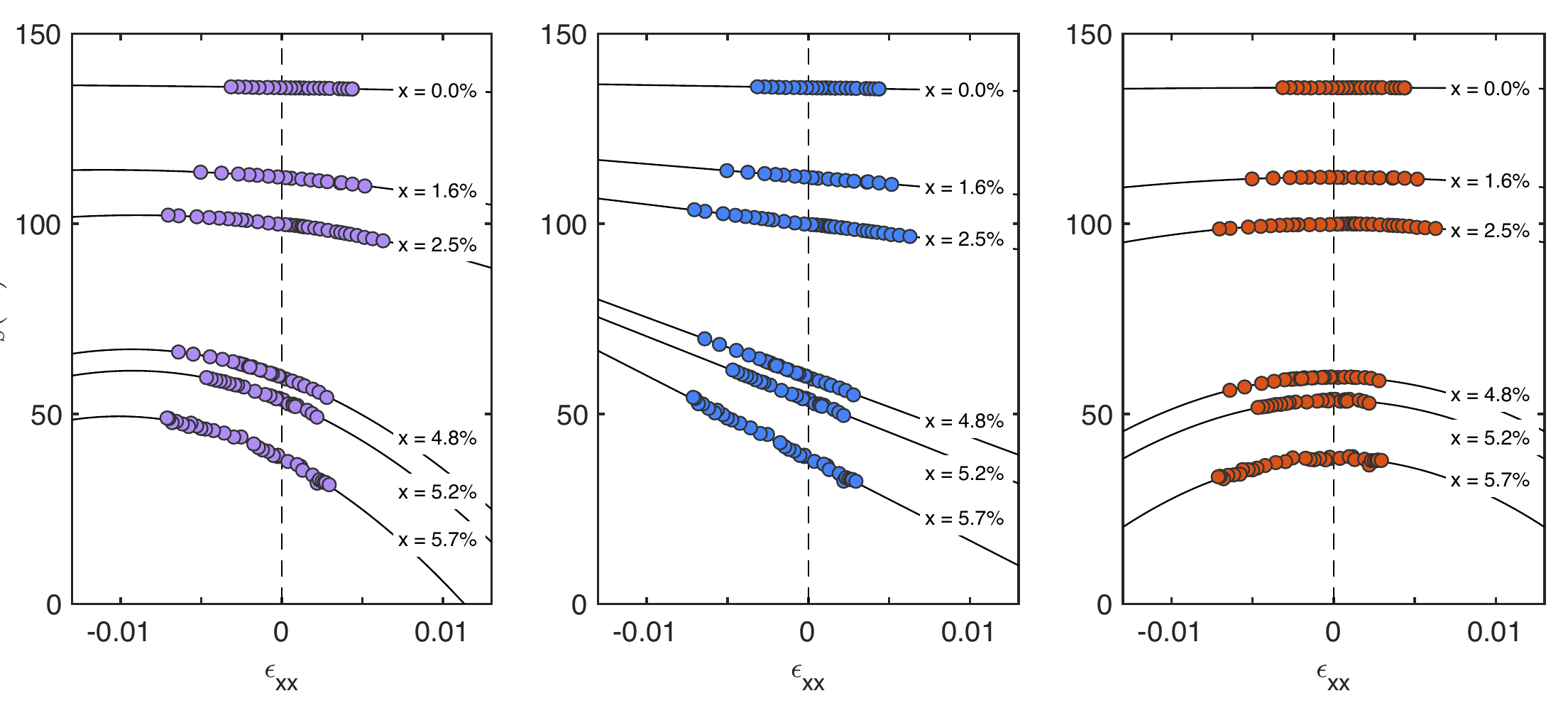}
  \caption{{\bf Response of $\mathbf{T_S}$ to applied uniaxial stress along [100] in various cobalt concentrations of $\mathbf{Ba(Fe_{1-x}Co_x)_2As_2}$.} (A) The purple circles represent the measured critical temperature $T_S$ as a function of the deduced strain $\epsilon_{xx}$. The black lines represent linear + quadratic fits to the data. As the cobalt concentration approaches the critical doping of $x_c \sim 6.7\%$, both the quadratic and linear coefficients increase. Panel (B) and (C) represent the linear and quadratic constituents due to $A_{1g}$ and $B_{1g}$, respectively.}
 \label{FigUni}
\end{figure*}

\section{\label{Apndx:Decomposing} Decomposing the variation of $T_S$ due to tuning strains}

In this section, we summarize our previous findings \cite{mitw}, explaining how the effect of each \emph{tuning strain} ($\epsilon_{A_{1g,1}}$, $\epsilon_{A_{1g,2}}$, and $\epsilon_{B_{1g}}$) on $T_S$ can be disentangled through the comparison of uniaxial stress and hydrostatic pressure experiments.

\subsection{Irreducible representation of strain}

When describing an arbitrary deformations caused by a uniform stress on a crystal, it is useful to express the strain as a linear combination of irreducible representations of the crystal. 
The $D_{4h}$ point group, appropriate for BaFe$_2$As$_2$ system, comprises of the following 6 irreducible representations: two symmetry preserving strain - $\epsilon_{A_{1g,1}} = (\epsilon_{xx} + \epsilon_{yy})/2$ and $\epsilon_{A_{1g,2}}=\epsilon_{zz}$, two C4 rotational symmetry breaking strain - $\epsilon_{B_{1g}} = (\epsilon_{xx} - \epsilon_{yy})/2$ and $\epsilon_{B_{2g}} = \epsilon_{xy}$, and two vertical shear strain - $\epsilon_{E_{g}} = (\epsilon_{xz}, \epsilon_{yz})$ components. 

In this work, we focus on the \emph{tuning strain} components - defined as strains that move the critical temperature the nematic phase transition without smearing the transition.
Symmetry preserving strains, $\epsilon_{A_{1g,1}}$, and $\epsilon_{A_{1g,2}}$, and the orthogonal antisymmetric strain, $\epsilon_{B_{1g}}$ are classified as tuning strains, while the longitudinal antisymmetric strain, $\epsilon_{B_{2g}}$, is not since it induces nematic order and smears the thermodynamic signatures of the phase transition. Further, vertical shear strain is not expected within our experiments and will be neglected in the discussion here.

\subsection{Disentangling the quadratic response}

The variation of $T_S$ due to the tuning strains ($\epsilon_{A_{1g,1}}$, $\epsilon_{A_{1g,2}}$, and $\epsilon_{B_{1g}}$) can be written as.
\begin{multline}
	T_S(\epsilon_{A_{1g,1}}, \epsilon_{A_{1g,2}},  \epsilon_{B_{1g}}) = T_0 + \sum_{i=1}^2 \lambda_{(A_{1g,i})}\epsilon_{A_{1g,i}} \\ +  \sum_{i\leq j =1}^2 \lambda_{(A_{1g,i}, A_{1g,j})}\epsilon_{A_{1g,i}} \epsilon_{A_{1g,j}} + \lambda_{(B_{1g},B_{1g})} \epsilon_{B_{1g}}^2 + ...
	\label{eq:Tsallstrain}
\end{multline}
where $T_0 = T_S(0,0,0)$ is the free standing (zero strain) critical temperature, $\lambda_{(a, b, ...)} \equiv \partial^nT_S/\partial \epsilon_a \partial \epsilon_b,...$. 
The coefficients $\lambda_{(B_{1g})}$ and  $\lambda_{(A_{1g,i},B_{1g})}$ are strictly zero due to symmetry constraints, and hence the corresponding terms are omitted in the above equation.

In hydrostatic pressure experiment, the magnitude of each irreducible strain can be calculated from the measured hydrostatic pressure and is linearly proportional to the pressure $P$ (i.e. $\epsilon_{A_{1g,i}} \propto P$).
Since hydrostatic pressure cannot induce antisymmetric strain, $\epsilon_{B_{1g}} = 0$, one can write the variation of $T_S$ as a function of hydrostatic pressure $P$ to be:
\begin{equation*}
	T_S(P) = T_0+AP+BP^2
\end{equation*}
Where $A$ is proportional to some combination of $\lambda_{(A_{1g,i})}$ and $B$ is proportional to some combination of  $\lambda_{(A_{1g,i}, A_{1g,j})}$. 

Fig.\ref{FigHyd} shows the result from a hydrostatic pressure experiments. Within the pressure range investigated, we observe only linear variations of $T_S$ as a function of pressure. This result suggests that all coefficients $\lambda_{(A_{1g,i}, A_{1g,j})}$ are zero. This reduces Eq.(\ref{eq:Tsallstrain}) to
\begin{multline*}
	T_S(\epsilon_{A_{1g,1}}, \epsilon_{A_{1g,2}},  \epsilon_{B_{1g}}) = T_0 + \sum_{i=1}^2 \lambda_{(A_{1g,i})}\epsilon_{A_{1g,i}} \\ +  \lambda_{(B_{1g},B_{1g})} \epsilon_{B_{1g}}^2 + ...
\end{multline*}
This attributes the linear variation of $T_S$ in uniaxial stress experiments to the response to symmetry preserving strain and the quadratic variation to orthogonal antisymmetric strain as illustrated in Fig.\ref{FigUni}.

\subsection{Variation in elastic moduli in differing sample dopings}

In the main text, the symmetry preserving strain is expressed as $\epsilon_{A_{1g}}$ without decomposing into the in-plane and out-of-plane components.
This $\epsilon_{A_{1g}}$ is a linear combination of $\epsilon_{A_{1g,1}}$ and $\epsilon_{A_{1g,2}}$ and the relative magnitude of the two strain modes depends on the applied uniaxial stress, and the elastic moduli of the samples.
If the elastic moduli change, the ratio of $\epsilon_{A_{1g,1}}$ and $\epsilon_{A_{1g,2}}$ will change which could in turn alter Eq.(4).
Here we will show that in the range of our experiment, the effect of changes in the elastic moduli is very small, such that the power law suggested in the Eq.(4) stays intact.

According to \cite{ElasticModuli}, the elastic tensors for $x=3.7\%$ and $x=6.0\%$ samples are
\begin{equation*}
C_{3.7\%}=
\left( \begin{array}{ccc}
93 & 11  & 11 \\ 
11  & 93  & 11 \\ 
11  & 11  & 88
\end{array} \right)
GPa,
\end{equation*}
and 
\begin{equation*}
C_{6.0\%}=
\left( \begin{array}{ccc}
112 & 27  & 27 \\ 
27  & 112  & 27 \\ 
27  & 27  & 84
\end{array} \right)
GPa
\end{equation*}
respectively. Note here that since $C_{13}$ cannot be measured accurately because of the crystal shape, $C_{13}$ is assumed to be equal to $C_{12}$ \cite{ElasticModuli} .

The relation between $\epsilon_{A_{1g,1}}$ and $\epsilon_{A_{1g,2}}$, and $\epsilon_{xx}$ is found to be $\epsilon_{A_{1g,1}} = \frac{1}{2}(1-\nu)\epsilon_{xx}$ and $\epsilon_{A_{1g,2}} = -\nu'\epsilon_{xx}$, where $\nu$ and $\nu'$ are in-plane and out-of-plane Poisson's ratio. From the elastic tensors, one obtain $\nu_{3.7\%} = 0.11$ and $\nu'_{3.7\%} = 0.27$ for $x=3.7\%$ cobalt concentration, and $\nu_{6.0\%} = 0.18$ and $\nu'_{6.0\%} = 0.26$ for $x=6.0\%$ cobalt concentration. This change in the Poisson ratios causes the prefactors $-\nu'$ and $\frac{1}{2}(1-\nu)$ to change by $-4\%$ and $-9\%$ from $3.7\%$ to $6.0\%$ cobalt concentration. For $\epsilon_{B_{1g}}=\frac{1}{2}(1+\nu)\epsilon_{xx}$, the change of the prefactor is $+6\%$.

The changes of the Poisson ratios enters in the coefficients $\alpha$ and $\beta$ in Eq.(2). 
However, the coefficients $\alpha$ and $\beta$ change by more than 100\% from $x=4.8\%$ to  $x=5.7\%$. 
Thus the effect of change in elastic moduli is negligible to wide extent.

Furthermore, these changes are unlikely to contribute to the power law dependence in temperature. Consider a power law in case of two tuning parameters
\begin{equation*}
	T_S = (T_0^{1/\theta}-a_1\epsilon_{A_{1g,1}}-a_2\epsilon_{A_{1g,2}})^\theta = (T_0^{1/\theta}-C\epsilon_{xx})^\theta
\end{equation*}
where $a_1$ and $a_2$ are constants and $C = \frac{1}{2}(1-\nu)a_1 - \nu'a_2$. The change in $C$ is at most -9\% as shown above.
Taylor expansion to leading order yields Eq.(4):
\begin{equation}
	\frac{T_S}{T_0} =  1 - \frac{C\theta}{T_0^{1/\theta}} \epsilon_{xx} +O(\epsilon_{xx}^2),
	\label{eq:Tsvsexx}
\end{equation}
Since C has absorbed the change in the Poisson ratios, one would expect it to change as a function of chemical substitution.
Given that there is no critical behavior in the elastic moduli \cite{ElasticModuli}, we assume that $C$ changes slowly and linearly in doping: $C = C_0 + k(x_c-x)$ where $x_c$ is the critical doping. Substitution into Eq.(\ref{eq:Tsvsexx}) gives
\begin{multline*}
	\frac{T_S}{T_0} =  1 - \frac{(C_0+k(x_c-x))\theta}{T_0^{1/\theta}} \epsilon_{xx} +O(\epsilon_{xx}^2) \\ =  1 - \left(\frac{C_0\theta}{T_0^{1/\theta}}+D \right) \epsilon_{xx} +O(\epsilon_{xx}^2)
\end{multline*}

where the last equality comes from the fact that $(x_c-x) \sim T^{1/\theta}$. This shows that the slowly varying elastic moduli do not affect the temperature dependence of the linear coefficient $\alpha/T_0$. 
The same result can analogously also be found for $B_{1g}$ strain.

\section{\label{Apndx:Scale} Scaling collapse of the critical temperatures}

\begin{figure}
	\includegraphics[width=1\linewidth]{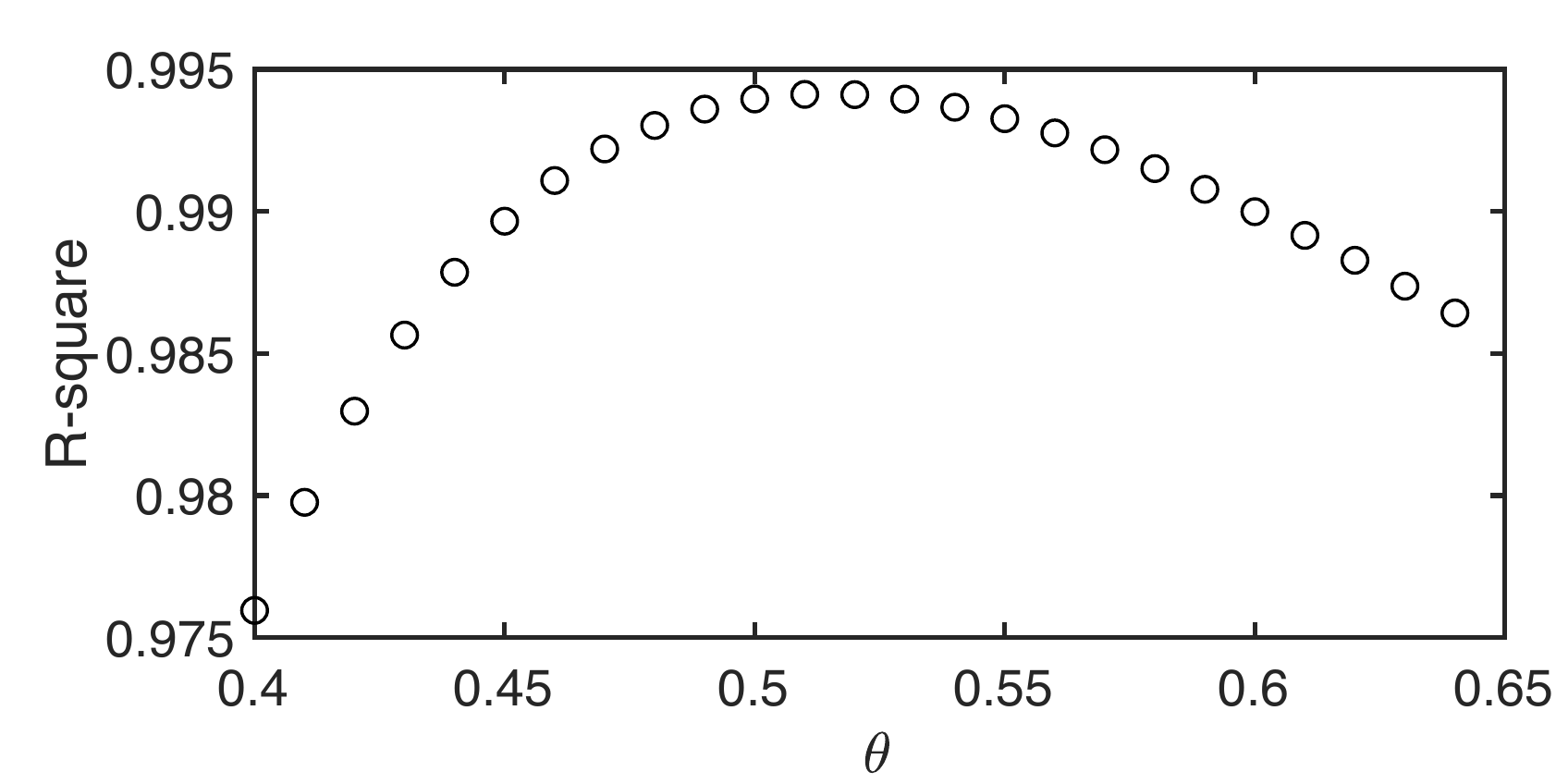}
	\caption{{\bf Theta scan.} The plot shows the R-square value of the least square fit of Eq.(\ref{eq:stevefit}) for various $\theta$ values ranging from $0.4-0.65$. $\theta = 0.52 \pm 0.02$ give the best fit with R-equare of 0.994.}
	\label{FigTheta}
\end{figure}

The variation of $T_S$ in the quantum critical regime can be written as 
\begin{equation}
	T_S(\delta) = |\delta|^\theta
	\label{eq:Tsdelta}
\end{equation}
Where $\delta$ is the magnitude of the "relevant" perturbation which is thus a linear combination of the various tuning parameters. In the context of our experiment, $\delta$ can be expressed as 
\begin{equation*}
	\delta = A_1 \epsilon_{A_{1g,1}} + A_2 \epsilon_{A_{1g,2}} + A_3 \epsilon_{B_{1g}}^2+A_4(x - x_c),
\end{equation*}
where $A_j$, for $j=1-4$, are constants. 
In quantum critical regime, $x-x_c \sim T_0^{1/\theta}$ and since $\epsilon_{A_{1g,1}}$,$\epsilon_{A_{1g,2}}$, and $\epsilon_{B_{1g}}$ are all proportional to $\epsilon_{xx}$, Eq.(\ref{eq:Tsdelta}) reduces to 
\begin{equation}
	T_S = |\delta|^\theta = (T_0^{1/\theta} +a\epsilon_{xx} + b\epsilon_{xx}^2)^\theta.
	\label{eq:Tsdeltaexx}
\end{equation}

\begin{figure}
	\includegraphics[width=1\linewidth]{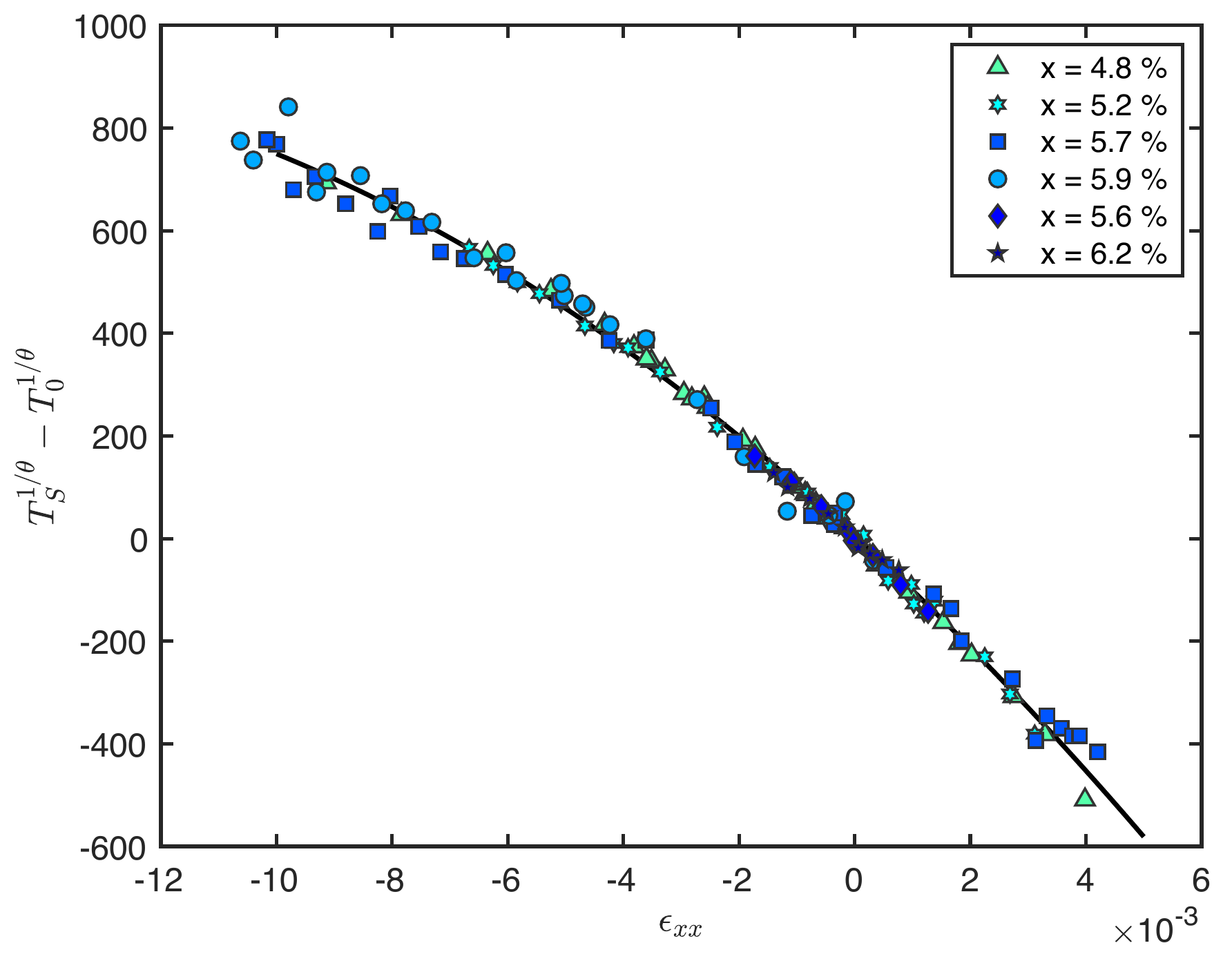}
	\caption{{\bf  Scaling collapse of critical temperature as a function of strain.} Symbols represent samples with differing dopings. It can be seen that the data collapse into a single curve. The quadratic fit described in Eq.(\ref{eq:stevefit}) is shown by the line with the best fit value of $\theta = 0.52 \pm 0.02$.}
	\label{FigFit}
\end{figure}

Note that there the constant $A_0$ vanishes since $T_0$ is defined as the critical temperature in an absence of strain. 
In order to fit to $a$,$b$, and $\theta$, we will rearrange Eq.(\ref{eq:Tsdeltaexx}) to 
\begin{equation}
		T_S^{1/\theta} - T_0^{1/\theta} =  a\epsilon_{xx} + b\epsilon_{xx}^2.
		\label{eq:stevefit}
\end{equation}
The best fit value of $\theta$ can be obtained from an R-square analysis of this quadratic fit in the quantum critical regime. 
Samples with a doping from $4.8\%$ to $6.2\%$ exhibit power law behavior of the critical temperature with $\theta = 0.56 \pm 0.07$ (Fig.\ref{FigResult}).
Using this information we scan $\theta$ from $0.4-0.6$ for these samples.
This is shown in Fig.\ref{FigTheta} where the best fit value is $\theta = 0.52 \pm 0.02$ (the uncertainty is obtained using Trust-Region optimization methods). 
Fig.\ref{FigFit} shows the quadratic fit with this value of $\theta$. 
All the data from the different samples collapse onto a single line.

\bibliographystyle{apsrev}
\bibliography{main.bib}

\begin{thebibliography}{39}
\bibitem[{Tsb()}]{Tsb1g}
\bibinfo{note}{In a tetragonal system, positive and negative $B_{1g}$ has the
  same effect to the material, therefore, $T_s$ can only be an even function of
  $B_{1g}$ strain - quadratic to leading order.}
\bibitem[{nuz()}]{nuz}
\bibinfo{note}{In cases where a scaling hypothesis holds, the critical
  temperature is set by an energy scale that is inversely proportional to the
  critical correlation time $\tau_C \sim (g_C-g)^{-\nu z}$ where $\nu$ is
  critical exponent for the correlation length, and $z$ is dynamical scaling
  exponent. In such case, $\theta = \nu z$ \cite{sachdev}, but we persist in
  writing the exponent simply as $\theta$ to preserve generality.}
\bibitem[{geb()}]{geb1gdquare}
\bibinfo{note}{Positive and negative $B_{1g}$ strain suppress the critical
  temperature equally, hence the square on $\epsilon_{B_{1g}}$}.
\end{thebibliography}
\end{document}